# Large anisotropic uniaxial pressure dependencies of $T_c$ in single crystalline Ba(Fe$_{0.92}$Co$_{0.08}$)$_2$As$_2$

Frédéric Hardy,[1] Peter Adelmann,[1] Hilbert v. Löhneysen,[1,2] Thomas Wolf,[1] and Christoph Meingast[1]

[1] Forschungszentrum Karlsruhe, Institut für Festkörperphysik, PO Box 3640, 76021 Karlsruhe

[2] Physikalisches Institut, Universität Karlsruhe, 76128 Karlsruhe.

Using high-resolution dilatometry, we study the thermodynamic response of the lattice parameters to superconducting order in a self-flux grown Ba(Fe$_{0.92}$Co$_{0.08}$)$_2$As$_2$ single crystal. The uniaxial pressure dependencies of the critical temperature of $T_c$, calculated using our thermal expansion and specific heat data via the Ehrenfest relation, are found to be quite large and very anisotropic ($dT_c/dp_a$ = 3.1(1) K/GPa and $dT_c/dp_c$ = - 7.0(2) K/GPa). Our results show that there is a strong coupling of the $c/a$ ratio to superconducting order, which demonstrates that $T_c$ is far from the optimal value. A surprising similarity with the uniaxial pressure effects in several other layered superconductors is discussed.

The recent discovery of superconductivity up to 26 K in a layered FeAs compound [1] has initiated a large research effort in these materials [2]. At the time of writing, the maximum values of the critical temperature, $T_c$, of around 55 K [3,4] have been found in the ReOFeAs (Re – rare earth) or "1111" structural family. Superconductivity also occurs in a simpler class of materials based on the BaFe$_2$As$_2$ parent compound ("122" structure) with $T_c$ values up to 38 K for K-doped 122 [5]. The 122 materials are particularly interesting for experimental studies, because fairly large flux-grown single crystals can readily be obtained [6,7]. Besides doping, pressure is an important tuning parameter in these systems, because the parent compounds actually become superconducting under moderate pressures of 0.5 to 5 GPa [8,9, 10]. The effect of hydrostatic pressure on $T_c$ has been studied by a number of groups in both the 1111 and 122 families [8,11-13]. Since the crystal structure of these new superconductors consists of conducting layers of $Fe_2As_2$, separated by other structural elements, one expects the pressure dependence of $T_c$ to be highly anisotropic, as observed in the cuprate superconductors [14- 17]. A large pressure dependence of $T_c$ at optimal doping indicates that $T_c$ can be further optimized by, for example, chemical pressure or expitaxial strain in thin films. The pressure derivatives of $T_c$ can also be useful for testing theories of superconductivity, especially if other pressure dependent quantities are determined as well [17,18].

In this Letter we study the anisotropic lattice response to superconducting order in a near optimally doped $Ba(Fe_{0.92}Co_{0.08})_2As_2$ single crystal using high-resolution thermal expansion measurements. We find clear anomalies in the thermal expansivities at $T_c$, which we use to calculate the uniaxial pressure dependencies of $T_c$ via the Ehrenfest relationship using specific heat data obtained on the same sample. We show that the pressure effect upon $T_c$ is very anisotropic, coupling strongly to the *c/a* ratio of lattice constants and largely cancels for hydrostatic pressure. Our results imply that $T_c$ in $Ba(Fe_{0.92}Co_{0.08})_2As_2$ is far from the optimal value. Much higher $T_c$ values could possibly be obtained by applying uniaxial strain (for example, epitaxial strain in thin films). Finally, we discuss surprising similarities of the present uniaxial pressure effects with those in other layered superconductors.

Co-doped Ba122 single crystals [7] were grown from self-flux in an alumina crucible. Prereacted FeAs and CoAs powders were mixed with Ba in the ratio Ba:Fe:Co:As = 1:2.67:0.33:3, placed into an $Al_2O_3$ crucible, which then was sealed in an evacuated $SiO_2$ ampoule. After heating to 650°C and then to 1170°C, with holding times of 5 hours, crystal growth took place during cooling down at a rate of 1.5 °C/h. At 1026 °C the ampoule was tilted to decant the remaining liquid flux from the crystals and subsequently pulled out of the furnace. Several crystals were investigated, and here we report on data for a crystal with dimensions of 1.8 x 1 x 0.25 mm$^3$. The composition was determined by energy dispersive x-ray spectroscopy to be Ba(Fe$_{0.92}$Co$_{0.08}$)$_2$As$_2$Co, implying that the crystal is close to optimal Co doping [19,20]. The resistivity, magnetization and specific heat were measured in a PPMS (Physical Properties Measurement System) from Quantum Design. The thermal expansion was measured in a home-made capacitance dilatometer. [21].

The superconducting transition at $T_c$ = 19.8 K is clearly seen in the field cooled (FC) and zero-field cooled (ZFC) magnetization curves in a field of 5 mT in Fig. 1 [22]. Fig. 2 shows the resistive transition in zero field, the midpoint of which (21.8 K) is somewhat higher than the magnetic transition. The resistance varies linearly with temperature from $T_c$ to about 125 K, as has been observed in other studies near the optimal $T_c$ value [19,20]. The resistivity ratio between $T_c$ and room temperature is about 4.

Fig. 3a shows the linear expansivities, $\alpha_i = 1/L_i \cdot dL_i/dT$, along the a- and c-axes from 5 K to 275 K. The thermal expansivity perpendicular to the Fe$_2$As$_2$ planes, $\alpha_c(T)$, is roughly three times larger than the in-plane expansivity, $\alpha_a(T)$, which is most likely due to phonon anharmonicity. Since there is usually a strong correlation between the thermal expansivity and the linear compressibility, our results imply that the compressibility is considerably larger along the *c*-axis. Clear second-order jumps in the thermal expansivity are seen in response to the onset of superconducting order at $T_c$ along both crystallographic directions (see Fig. 3b). This is expected to be a purely electronic effect, related to the volume dependence of the superconducting condensation energy. We find $\Delta\alpha_a$ = 1.4(1) · 10$^{-6}$ K$^{-1}$ and $\Delta\alpha_c$ = -3.2(1) · 10$^{-6}$ K$^{-1}$ for the jumps in the expansivities along the a- and c-axes, respectively.

The uniaxial pressure dependencies of $T_c$, $dT_c/dp_i$, are given by the Ehrenfest relationship,

$$\frac{dT_c}{dp_i} = \frac{\Delta\alpha_i V_m}{\Delta C_p / T_c}, \qquad (1)$$

where $\Delta\alpha_i$ is the jump in the expansivity along the i direction, $V_m = 61.3$ cm$^3$/mole [7] is the molar volume, and $\Delta C_p$ is the specific-heat jump. Fig. 4 shows the heat capacity data after the subtraction of a reasonable phonon background [23]. The data show a clear jump with $\Delta C_p/T_c = 28$ mJ/moleK$^2$. There is a non-zero residual value of the linear specific-heat coefficient of about 7 mJ/moleK$^2$ at low $T$ in the data; comparable in value to other studies [7] and possibly indicating a small amount of metallic FeAs flux remaining on the crystal. Using our measured thermal expansion and specific-heat jumps we find $dT_c/dp_a = 3.1(1)$ K/GPa and $dT_c/dp_c = -7.0(2)$ K/GPa.

Our results show that the pressure dependence of $T_c$ in Ba(Fe$_{0.88}$Co$_{0.12}$)$_2$As$_2$ is very anisotropic and quite large in magnitude. We note that under hydrostatic pressure conditions, the uniaxial pressure dependencies largely cancel and result in a much smaller negative pressure dependence of $T_c$; the volume pressure effect $dT_c/dp_{vol.}$ is given by $dT_c/dp_{vol.} = 2\, dT_c/dp_a + dT_c/dp_c = -0.9(3)$ K/GPa. For the FeAs superconductors, as in cuprates [17, 24], the pressure effect is composed of two terms. The first term is related to the pressure-induced doping of the system, and the second one is an 'intrinsic' pressure effect, which changes the strength of the pairing mechanism, and thus the maximum $T_c$ value. At optimal doping, the initial pressure dependence of $T_c$ probes only the intrinsic effect [17,24]. The strong sensitivity of $T_c$ to anisotropic strain near optimal doping observed in our data indicates that the intrinsic effect is quite sizable, and that $T_c$ is far from the optimal value in Ba(Fe$_{0.88}$Co$_{0.12}$)$_2$As$_2$; Our data imply that $T_c$ can be increased significantly by increasing the $c/a$ ratio, and here it is interesting to note that Co doping decreases $c/a$, whereas K doping increases $c/a$ [5,7]. The difference in the maximal $T_c$ values of Co- and K–doped Ba122 ($T_c =$ 22 K and 39 K, respectively), may, thus, largely be due to a chemical pressure effect upon *c/a*. Values for the elastic constants are, however, needed for a quantitative determination of this effect. Epitaxial strain could also be used to change *c/a* and, thus, $T_c$, as has been shown for the single-layer cuprate superconductor La-Sr-Cu-O, for which $T_c$ was actually doubled by epitaxial strain [25]. The physical origin of the large observed sensitivity of $T_c$ to small structural changes is unclear, but is most likely related to the strong sensitivity of the calculated electronic band structure to small structural changes [26-28]. Besides the Fe-As distance, the angle of the As-Fe-As bonds appears to strongly correlate with $T_c$ [29]. Our results provide a strong constraint for theories of superconductivity in these materials. Further thermal expansion experiments are in progress to correlate the present results with the uniaxial pressure dependence of the Sommerfeld coefficient.

In summary, the response of the crystal lattice to superconducting order in a close-to-optimally-doped $Ba(Fe,Co)_2As_2$ single crystal was studied using high-resolution thermal expansion measurements. The uniaxial pressure dependencies of $T_c$ were determined and found to be very anisotropic and quite large in magnitude; increasing the *c/a* ratio is expected to increase $T_c$. These effects largely cancel for hydrostatic pressure. Interestingly, several other layered superconductors show a strong dependence of $T_c$ on *c/a*, as well, notably, the heavy-fermion systems $PuMGa_5$ (M = Co; Rh) and $CeMIn_5$ (M = Co; Rh; Ir) [30] and the cuprate superconductors La-Sr-Cu-O [15], Bi2212 [16] and Hg1201 [31]. Although these systems exhibit important differences among each other, notably the $T_c$ values, they have in common the quasi-two-dimensional electronic structure. It is intriguing that the $T_c$ values increases with increasing *c/a* for all of these systems. Possibly, this points to a general importance of low dimensionality in these unconventional superconductors.

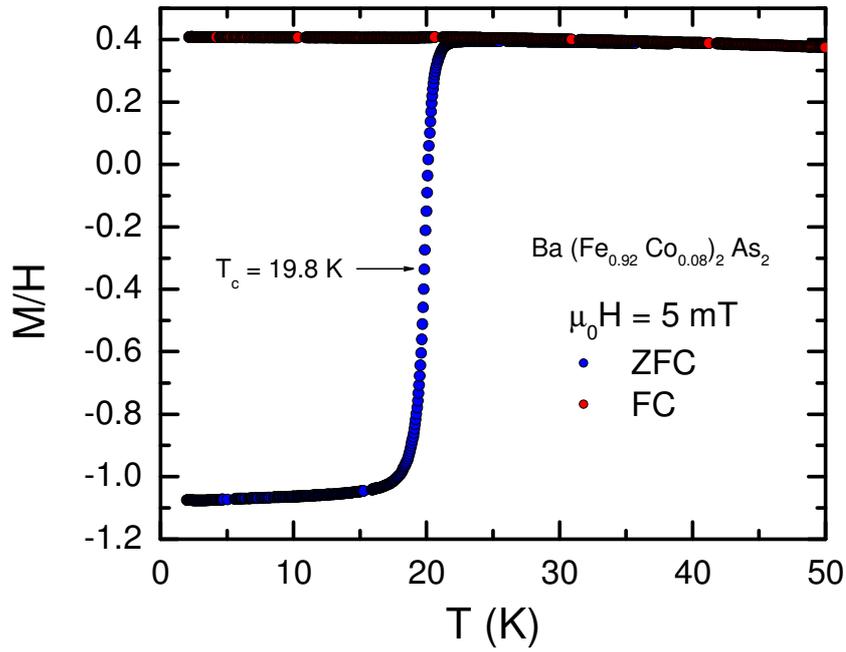

Fig. 1. Field cooled (FC) and zero-field-cooled (ZFC) magnetization in a field of 5 mT.

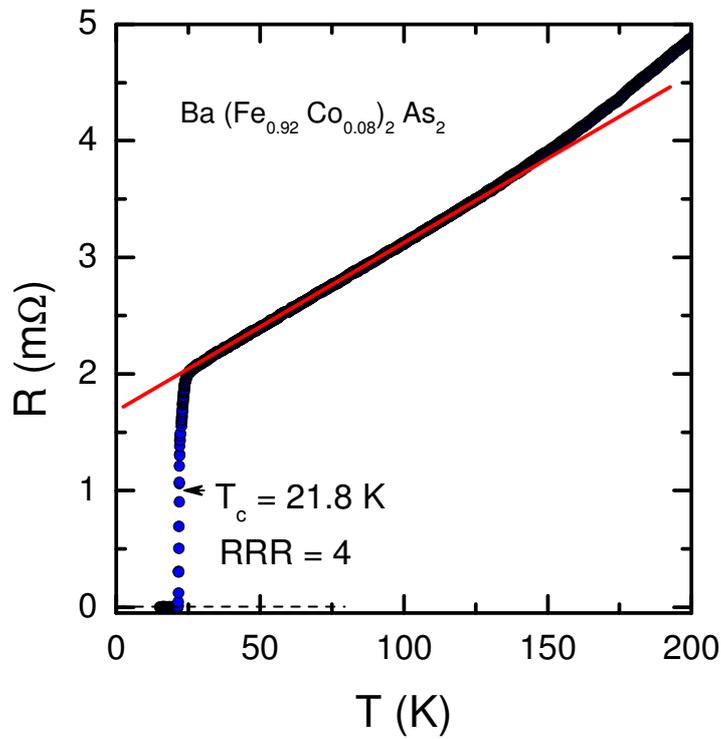

Fig. 2. Resistance versus temperature.

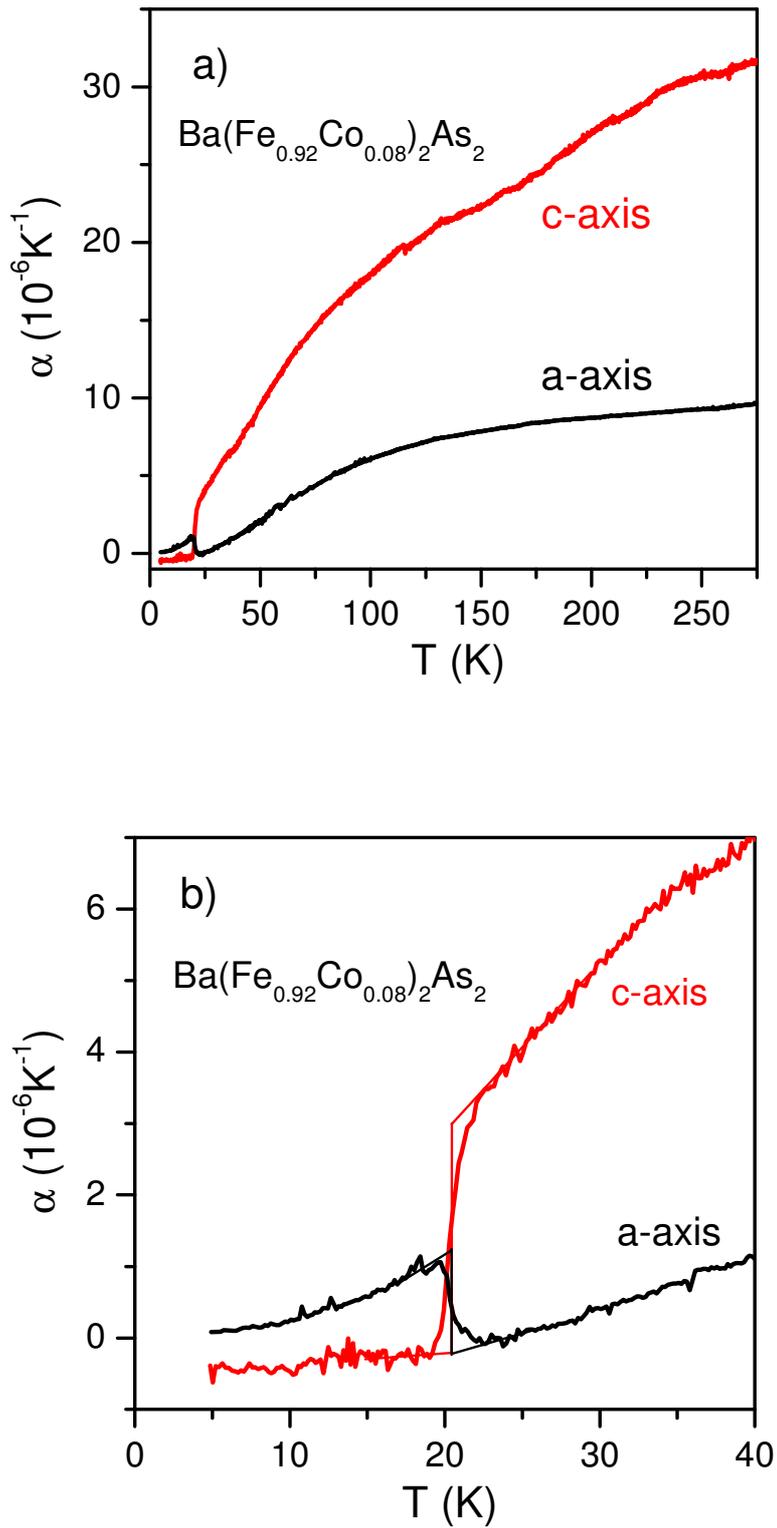

Fig. 3. Linear thermal expansivity versus temperature a) between 5 K and 275 K and b) around $T_c$.

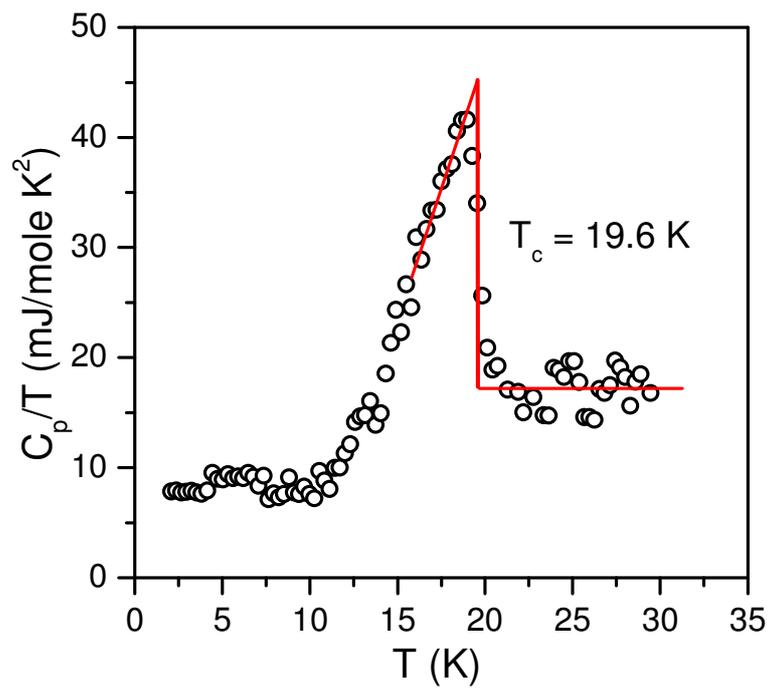

Fig. 4. Specific-heat anomaly at $T_c$ obtained by subtracting a phonon background (see text for details).